\documentclass[prd,aps,superscriptaddress,unsortedaddress,twocolumn,%
  preprintnumbers,amsmath,amssymb]{revtex4-2}
\usepackage{graphicx}
\usepackage[breaklinks,colorlinks=true]{hyperref}
\usepackage{slashed}

\newcommand{\tr}{\mathop{\rm tr}}
\newcommand{\LMS}{\Lambda_{\overline{\text{MS}}}}
\newcommand{\nB}{n_{\text{B}}}
\newcommand{\bLambda}{\bar{\Lambda}}

\newcommand{\Nc}{N_{\text{c}}}
\newcommand{\Nf}{N_{\text{f}}}
\newcommand{\Ms}{M_{\text{s}}}
\newcommand{\bk}{\boldsymbol{k}}
\newcommand{\mD}{m_{\text{D}}}
\newcommand{\mqf}{m_{\text{q}f}}
\newcommand{\PHDL}{P_{\text{HDLpt}}}

\newcommand{\Cg}{C_{\text{g}}}
\newcommand{\Cq}{C_{\text{q}}}
\newcommand{\Dq}{D_{\text{q}}}

\newcommand{\tomega}{\tilde{\omega}}
\newcommand{\bomega}{\bar{\omega}}
\newcommand{\tTK}{\widetilde{\mathcal{T}}}
\newcommand{\muB}{\mu_{\text{B}}}
\DeclareMathOperator*{\SumInt}{%
\mathchoice%
  {\ooalign{$\displaystyle\sum$\cr\hidewidth$\displaystyle\int$\hidewidth\cr}}
  {\ooalign{\raisebox{.14\height}{\scalebox{.7}{$\textstyle\sum$}}\cr\hidewidth$\textstyle\int$\hidewidth\cr}}
  {\ooalign{\raisebox{.2\height}{\scalebox{.6}{$\scriptstyle\sum$}}\cr$\scriptstyle\int$\cr}}
  {\ooalign{\raisebox{.2\height}{\scalebox{.6}{$\scriptstyle\sum$}}\cr$\scriptstyle\int$\cr}}
}
\newcommand{\Disc}{\mathop{\mathrm{Disc}}}
\newcommand{\qzero}[2]{\ln\left(\frac{#1 + #2}{#1 - #2}\right)}
\renewcommand{\Im}{\mathop{\mathrm{Im}}}
\renewcommand{\Re}{\mathop{\mathrm{Re}}}

\begin{document}
\title{Equation of state of cold and dense QCD matter in resummed perturbation theory}
\author{Yuki~Fujimoto}
\author{Kenji~Fukushima}
\affiliation{Department of Physics, The University of Tokyo, %
  7-3-1 Hongo, Bunkyo-ku, Tokyo 113-0033, Japan}

\begin{abstract}
  We discuss the Hard Dense Loop resummation at finite quark mass
  and evaluate the equation of state (EoS) of cold and dense QCD matter
  in $\beta$ equilibrium.  The resummation in the quark sector has an
  effect of lowering the baryon number density and the EoS turns out
  to have much smaller uncertainty than the perturbative QCD
  estimate.  Our numerical results favor smooth matching between the
  EoS from the resummed QCD calculation at high density and the
  extrapolated EoS from the nuclear matter density region.  We also
  point out that the speed of sound in our EoS slightly exceeds the
  conformal limit.
\end{abstract}
\maketitle

\section{Introduction}
A reliable estimate of the equation of state (EoS) of cold matter at
high baryon density is a vital challenge in theoretical nuclear
physics.  In various circumstances such as the neutron star cores, the
neutron star mergers emitting gravitational waves, the supernova
explosion, and the heavy-ion collisions to scan over the phase diagram
of matter made out of quarks and gluons (see
Ref.~\cite{Fukushima:2020yzx} for a review on the present status and
the future direction of the heavy-ion collision), the EoS is an
indispensable input for theoretical studies.  Conversely, experimental
data available from these extreme environments provide us with useful
constraints on possible EoSs, so that some theoretical scenarios can
be excluded/accepted.  The most well-known and successful example
along these lines is the establishment of two-solar-mass neutron
stars~\cite{Demorest:2010bx, *Fonseca:2016tux,*Antoniadis:2013pzd,*Cromartie:2019kug},
which disfavors scenarios leading to soft EoS; namely, it is unlikely
for dense matter to accommodate a strong first-order phase
transition~\cite{Alford:2015dpa} nor condensations of exotic degrees
of freedom.

The most advanced first-principles approach from the fundamental
theory of the strong interaction, i.e., quantum chromodynamics (QCD)
is the lattice Monte-Carlo simulation, but the notorious sign problem
ruins the importance sampling algorithm for matter at finite baryon
density.  Still, in parameter space where the lattice-QCD simulation
is at work, the validity of alternative theoretical approaches has
been tested.  In particular, the Hard Thermal Loop perturbation theory
(HTLpt) is the most promising resummation
scheme~\cite{Andersen:1999fw,*Andersen:1999sf,Andersen:1999va,Haque:2014rua,
  Mogliacci:2013mca,Ghiglieri:2020dpq}
that confronts the lattice-QCD results at high temperature $T$.  The
purpose of this work is to quantify the resummation effects on the
EoS of cold and dense quark matter at high baryon density $\nB$ or the
energy density $\varepsilon$.

To sharpen novelties in our work, let us briefly summarize what has been
understood so far.  Since the seminal works of
Refs.~\cite{Freedman:1976xs,*Freedman:1976dm,*Freedman:1976ub,Baluni:1977ms},
we had to wait for about three decades until the perturbative QCD
(pQCD) EoS was augmented with the strange quark mass $\Ms\neq0$ and
applied to the neutron star phenomenology~\cite{Fraga:2004gz,
  Kurkela:2009gj}, where they found that the strange mass effect is
crucial.
The obstacle in utilizing the pQCD EoS in neutron star physics was
found to be too large scale variation uncertainty
in the intermediate density region (i.e.,
denser than the nuclear terrain but not dense enough to justify pQCD)
and the theoretical efforts are progressing toward further
higher-order calculations~\cite{Gorda:2018gpy,
  Gorda:2021znl,*Gorda:2021kme} with hope for better convergence
(see also for Refs.~\cite{Kneur:2019tao, Fernandez:2021jfr} for an
alternative approach based on the renormalization group optimization
method).

From the success of HTLpt at high $T$, it is a natural anticipation
that the same machinery of resummation would cure the convergence
problem at high baryon density or large quark chemical potential $\mu$
as well, which may reduce the scale variation uncertainty.
Indeed, the parallelism between the high $T$ and high $\mu$ cases has
been established based on the transport equation approach in
Ref.~\cite{Manuel:1995td}; the high-density counterparts of HTLs are
called Hard Dense Loops (HDLs).
As long as a resummation prescription in the quark sector is
concerned, more simply, we can just take the $T\to0$ limit of HTLpt to
introduce ``HDLpt'' as considered in Ref.~\cite{Baier:1999db} (see also
Ref.~\cite{Andersen:1999va}, and we note that the term
  ``HDLpt'' was first introduced in Ref.~\cite{Andersen:2002jz}).
The HTL approximation usually neglects the bare quark
mass and only the screening masses of quarks enter expressions used in
Refs.~\cite{Baier:1999db, Andersen:1999va}.  Later on, extensive
discussions about the EoS and the quark star properties have been
addressed in Ref.~\cite{Andersen:2002jz}.  As seen in Fig.~2 of
Ref.~\cite{Andersen:2002jz}, however, the HDLpt hardly remedies the
convergence problem associated with uncertainty of the scale
$\bLambda=\mu-4\mu$ in the running coupling constant
$\alpha_s(\bLambda)$.  In the present
work, as in Ref.~\cite{Kurkela:2009gj}, we will employ the two-loop formula;
$\alpha_s(\bLambda)=[1- 2(\beta_1/\beta_0^2)
\ln^2(\bLambda^2/\LMS^2)/\ln(\bLambda^2/\LMS^2)]\,
4\pi/[\beta_0\ln(\bLambda^2/\LMS^2)]$,
where
$\beta_0\equiv (11\Nc-2\Nf)/3$,
$\beta_1\equiv (17/3)\Nc^2 - \Nf (\Nc^2-1)/(2\Nc) - (5/3)\Nf\Nc$,
and we will take
$\LMS=378\,\text{MeV}$ throughout, following
Ref.~\cite{Kurkela:2009gj}.
Previously, the absence of the bare quark mass significantly simplified
technicalities as well as the realization of the $\beta$ equilibrium.
With equal amount of $u$, $d$, and $s$ quarks (that is automatically
the case if their masses are all neglected), the electric charge
neutrality follows as it is.  For quantitative descriptions of the
neutron star phenomenology, however,  we need to take account of the
strange quark mass and solve the $\beta$ equilibrium condition.

There seems to be a long way left, but the phenomenological analyses
are in need of the QCD-based EoS usable for the neutron star
observables.  In fact, on top of extrapolated EoSs from the nuclear
side, the Bayesian analysis has been recognized as a powerful
instrument for the inference analysis to identify the most likely EoS
based on the observational
data~\cite{Ozel:2010fw,Steiner:2010fz,Alvarez-Castillo:2016oln}
(see Ref.~\cite{Ozel:2016oaf} for a review).  Recently, the Machine
Learning technique has been also advocated as a complementary method
to infer the
EoS~\cite{Fujimoto:2017cdo,Fujimoto:2019hxv,Fujimoto:2021zas}.   It
would be of utmost importance to make a direct comparison of the
inferred EoS candidates and the QCD-based estimates.  To this end, we
are urged to reduce uncertainty and widen the validity region of the
pQCD or HDLpt calculations.

In this work we will report the first successful attempt to
construct an EoS with smaller uncertainty from the HDLpt framework
incorporating the strange quark mass effect.  From the technical point
of view, we adopt the resummation schemes in the gluon sector as
prescribed in Ref.~\cite{Andersen:1999fw,*Andersen:1999sf} and in the
quark sector as in Ref.~\cite{Baier:1999db} with our own extension to
cope with the strange quark mass.  Our expressions are given in the
form of exact integrations without any expansion in terms of the
screening mass as in Ref.~\cite{Mogliacci:2013mca}.
This paper is organized as follows: In Sec.~\ref{sec:central}, we
present our central results, namely the reduction of the scale
dependent uncertainty in the perturbative calculation with
resummation.
Sec.~\ref{sec:formulation} and Sec.~\ref{sec:details} show the
calculation of the EoS within the HDLpt.
In Sec.~\ref{sec:discussions}, we show the numerical results on the
speed of sound, and we take into account the order $\alpha_s$
correction.
Finally, Sec.~\ref{sec:summary} summarizes this paper.

\section{Central results}
\label{sec:central}

\begin{figure}
  \includegraphics[width=\columnwidth]{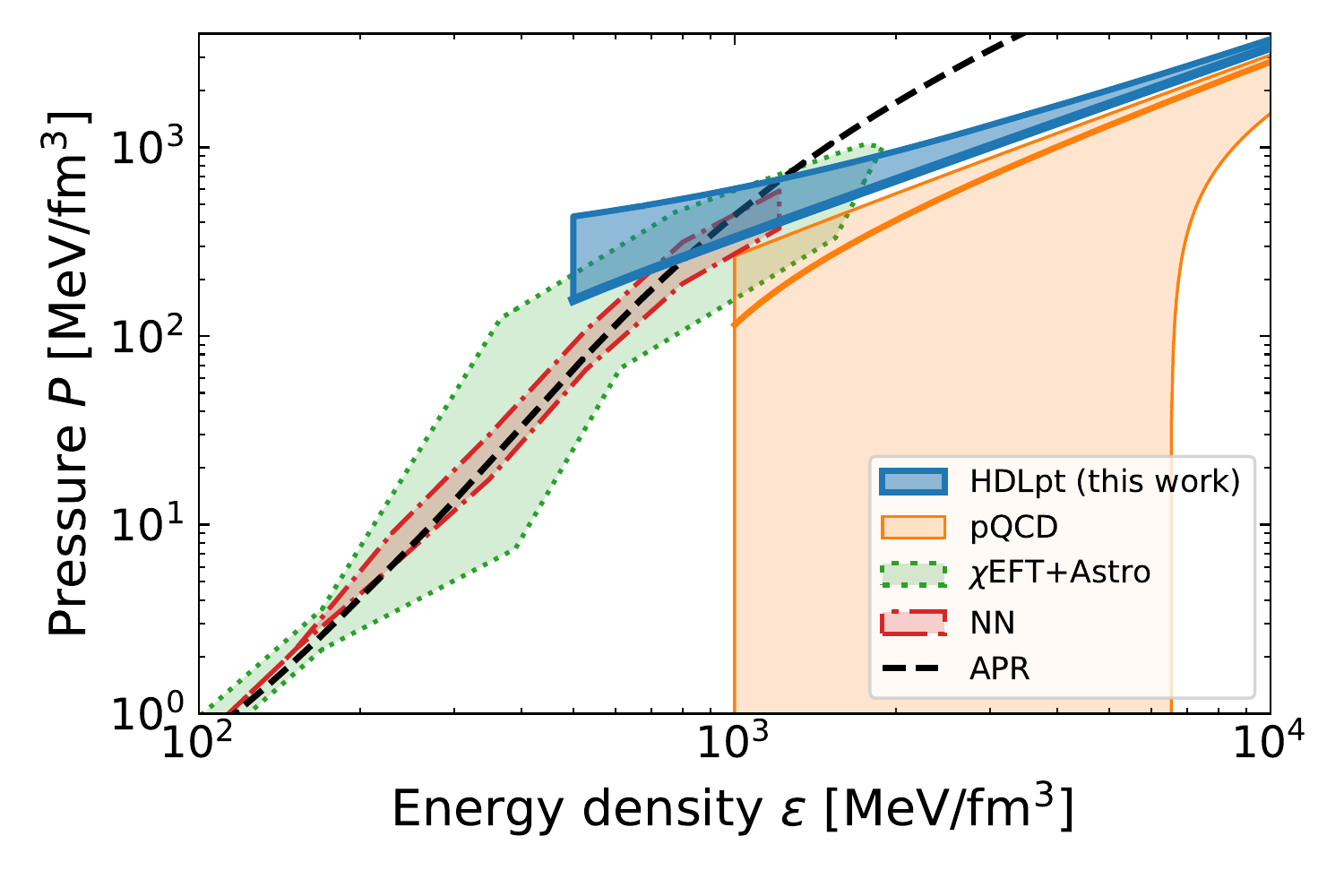}
  \caption{Comparison of the EoS in this work (HDLpt) and other
    EoSs.  The blue and the orange bands represent our results and
    the preceding results from Refs.~\cite{Kurkela:2009gj, Fraga:2013qra},
    respectively, with $\bLambda=\mu-4\mu$.  The green band is from the
    $\chi$EFT~\cite{Hebeler:2010jx}.  The red band shows the EoS
    inferred from the Neural Networks in the machine learning analysis
    of the neutron star observation~\cite{Fujimoto:2019hxv}.  The
    dashed black line is the APR EoS extrapolated from the nuclear
    side~\cite{Akmal:1998cf}.}
  \label{fig:pe3}
\end{figure}

Since technical details are cumbersome, we shall first present our
central results in Fig.~\ref{fig:pe3} and then proceed to technical
details later.  Not to make the comparison on the figure too busy, we
chose only a few representative EoSs from the nuclear side; namely,
the EoS extrapolated from the chiral Effective Field Theory
($\chi$EFT) calculation~\cite{Hebeler:2010jx} by the green band, the
Neural Network output in the machine learning
analysis~\cite{Fujimoto:2019hxv} by the red band, and the
Akmal-Pandharipande-Ravenhall (APR) EoS~\cite{Akmal:1998cf} shown by
the dashed line.

The orange band in the region,
$\varepsilon > 10^3\,\mathrm{MeV}/\mathrm{fm}^3$, represents the
results from pQCD~\cite{Kurkela:2009gj} for which we utilize the
concise formula as given in Ref.~\cite{Fraga:2013qra}.
Higher-order corrections could be added, but the uncertainty band is
not much changed from Ref.~\cite{Kurkela:2009gj}.  The uncertainty
band width abruptly diverges, from which it has been said that pQCD is
reliable only at extreme high densities far from reality.  At a
glance, indeed, we should understand how difficult it is to make a
robust interpolation between the nuclear and the pQCD EoSs.  Now, a
surprise comes from a blue narrow band that represents results from
our HDLpt calculations.  The uncertainty band is drastically reduced
and the HDLpt EoS appears to be merged into the nuclear EoSs smoothly
in the intermediate density region.  It should be noted that the APR
EoS overshoots ours, but this is due to a well-known flaw in the APR
EoS, i.e., superluminal speed of sound which violates causality.

\begin{figure*}
  \includegraphics[width=1.99\columnwidth]{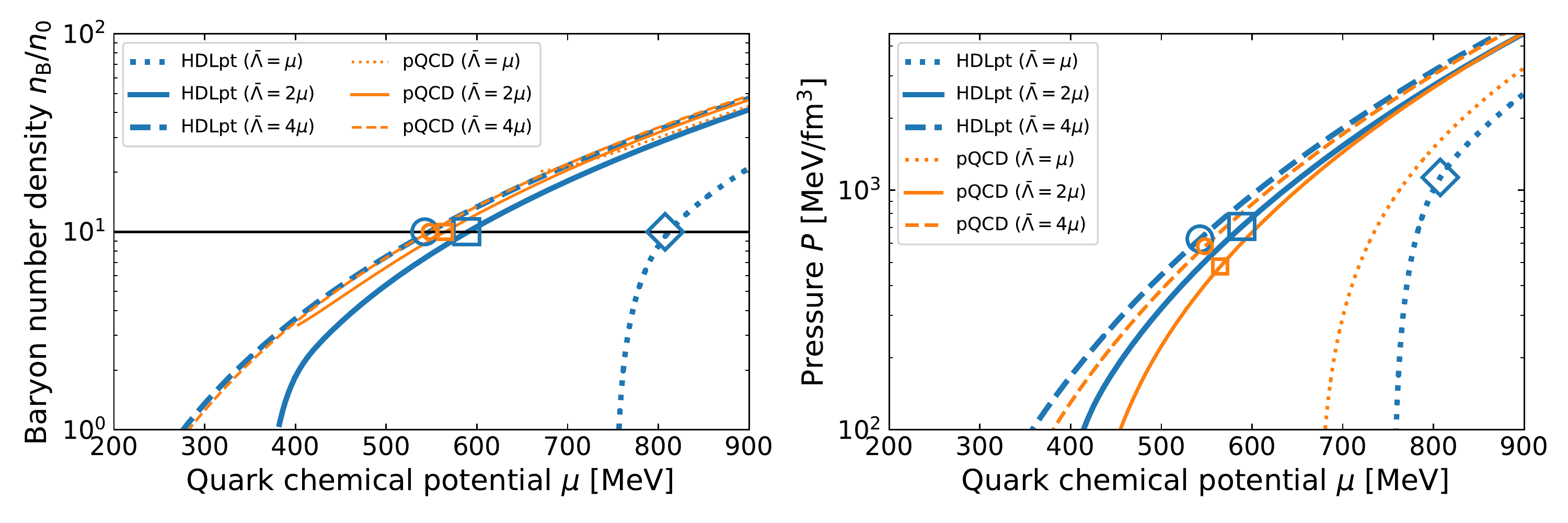}
  \caption{Baryon number density (left) and pressure (right)
    as functions of the quark chemical potential.  In the figure pQCD
    refers to the results from Refs.~\cite{Kurkela:2009gj,
      Fraga:2013qra} and HDLpt to our results.}
  \label{fig:nmu}
\end{figure*}

One may wonder what causes such a drastic difference on
Fig.~\ref{fig:pe3}.  We can qualitatively understand this from
Fig.~\ref{fig:nmu} (left) in which the baryon number density $\nB$ as a
function of the quark chemical potential $\mu$ is plotted.  Because
the HDLpt sums the quark loops up, $\nB$ is the most sensitive
quantity affected by the resummation in the quark sector.  It is an
interesting and reasonable observation that $\nB$ is suppressed at
fixed $\mu$ after the resummation:  thermodynamic quantities are
dominated by quark quasi-particles, and in HDLpt, quark excitations
are more screened by self-energy insertions, as compared to pQCD
treatments.  Therefore, on Fig.~\ref{fig:pe3}, the corresponding $\mu$
for a given $\varepsilon$ becomes larger, and the corresponding
running coupling $\alpha_s(\bLambda=\xi\mu)$, where $\xi=1,2,4$, is
smaller.  This qualitative argument partially accounts for the
reduction of the uncertainty band, but not fully yet.  As
  shown in Fig.~\ref{fig:nmu}, if we plot the pressure $P$, the
baryon number density $\nB$, and the energy density $\varepsilon$ as
functions of $\mu$, respectively, the uncertainty bands are wider than
Fig.~\ref{fig:pe3}.
In Fig.~\ref{fig:nmu} (left), we overlay a horizontal line at
$\nB=10\,n_0$ to find the values of corresponding $\mu$ for different
$\bLambda$.  The values of $P$ at these $\mu$'s are shown in
Fig.~\ref{fig:nmu} (right) with the same markers.  Importantly, the
marker for $P (\bLambda=\mu$) is out of the plot range.
Owing to the suppression in $\nB$ leads to the situation that
$P(\varepsilon)$ with $\bLambda=\mu$ and that with
$\bLambda=4\mu$ happen to stay close, which narrows the uncertainty
band on Fig.~\ref{fig:pe3}.
There might be a deep reason (e.g., scaling properties) for this
behavior, and further investigations are in progress.

For the astrophysical application, we need $P(\varepsilon)$ or
$P(\nB)$ rather than $P(\mu)$.
The condition that $P(\varepsilon; \bLambda)$ is insensitive to the scale
$\bLambda$ is $dP(\varepsilon; \bLambda)/d\bLambda = 0$, i.e., 
\begin{equation}
  \label{eq:conditione}
  \frac{\partial P(\muB;\bLambda)}{\partial \bLambda}
  - c_s^2 \frac{\partial \varepsilon(\muB; \bLambda)}{\partial \bLambda} = 0\,,
\end{equation}
where $\mu_{\rm B} = 3\mu$ is the baryochemical potential.
Substituting the thermodynamic relation $\varepsilon = -P + \muB \nB$
this relation reduces to
\begin{equation}
  \label{eq:conditionnB}
  (1+c_s^2)\frac{\partial P(\muB;\bLambda)}{\partial \bLambda}
  - c_s^2 \muB \frac{\partial \nB(\muB; \bLambda)}{\partial \bLambda} = 0\,.
\end{equation}

In the conventional argument, the reduction of the first terms in
Eqs.~\eqref{eq:conditione} and \eqref{eq:conditionnB} has been
the central issue, but we point out that $\partial P / \partial
\bLambda=0$ is only a sufficient condition for
Eqs.~\eqref{eq:conditione} and \eqref{eq:conditionnB}.
Albeit $\partial P / \partial \bLambda \neq 0$, the inclusion of the
latter term can cancel the scale-dependence; Fig.~\ref{fig:nmu} is the
concrete realization of such cancellation.


\section{Formulation}
\label{sec:formulation}
Let us explain the formulae and procedures to obtain our results in
Fig.~\ref{fig:pe3}.
Dense matter in the neutron star reaches the $\beta$ equilibrium;
$d\leftrightarrows u+e^-+\bar{\nu}_e$ and
$s\leftrightarrows u+e^-+\bar{\nu}_e$ indicating the relations between
quark chemical potentials as
$\mu_u=\mu+\frac{2}{3}\mu_Q$ and
$\mu_d=\mu_s=\mu-\frac{1}{3}\mu_Q$ where $\mu_Q$ is the electric
chemical potential.  Since electrons are negatively charged,
$\mu_e=-\mu_Q$, and we can fix $\mu_Q$ from the charge neutrality,
i.e., $n_Q-n_e=0$ with $n_Q=\partial P/\partial\mu_Q$ and
$n_e=\mu_e^3/(3\pi^2)$ neglecting the electron mass.

Since the most crucial extension in this work is the inclusion of the
bare quark mass, we will write down the explicit expressions in the
quark sector.  In our notation for flavor-$f$ quarks the bare mass is
$M_f$ and the screening mass is $\mqf$.  The bare mass should be scale
dependent as
\begin{equation}
  M_f(\bLambda) = M_f(2\text{GeV})\biggl[ \frac{\alpha_s(\bLambda)}
  {\alpha_s(2\text{GeV})} \biggr]^{\gamma_0/\beta_0}
  \frac{1+\mathcal{A}(\bLambda)}{1+\mathcal{A}(2\text{GeV})}\,.
\end{equation}
Here, $\beta_0$ was already introduced when $\alpha_s(\bLambda)$
appeared before, and $\gamma_0\equiv 3(\Nc^2-1)/(2\Nc)$.  The two-loop
corrections appear in
$\mathcal{A}(\bLambda)\equiv A_1(\alpha_s(\bLambda)/\pi)+
\frac{A_1^2+A_2}{2}(\alpha_s(\bLambda)/\pi)^2$ with
$A_1\equiv -\beta_1\gamma_0/(2\beta^2)+\gamma_1/(4\beta_0)$
and
$A_2\equiv \gamma_0/(4\beta_0^2)(\beta_1^2/\beta_0-\beta_2)
-\beta_1\gamma_1/(8\beta_0^2)+\gamma_2/(16\beta_0)$.  For $\beta_2$,
$\gamma_1$, and $\gamma_2$, the general expressions are complicated,
and we refer to numerical values, $\beta_2=3863/24$,
$\gamma_1=182/3$, and
$\gamma_2=8885/9-160\zeta(3)\approx 794.9$ for $\Nc=\Nf=3$.  Readers
can consult Eq.~(8) of Ref.~\cite{Kurkela:2009gj} for the complete
expressions.

In the $T\to 0$ limit the HDLpt pressure, $\PHDL$, is given by the
gluon loop and the quark loop with the self-energy insertions; namely,
\begin{equation}
  \label{eq:PHDL}
  \PHDL = (\Nc^2-1)P_{\text{g}} + \Nc\sum_{f=u,d,s} P_{\text{q},f}
  + \Delta P_{\text{g},\text{q}}\,, 
\end{equation}
where $\Delta P_{\text{g}}$ and $\Delta P_{\text{q}}$ subtract the
ultraviolet divergences.  The gluon part with an appropriate
subtraction by $\Delta P_{\text{g}}\propto 1/\epsilon$ (where the
spatial dimensions are $d=3-2\epsilon$ in the dimensional
regularization) is
\begin{equation}
  P_{\text{g}} = \frac{\mD^4}{64\pi^2} \biggl(
  \ln\frac{\bLambda}{\mD} + \Cg \biggr)\,. 
\end{equation}
A constant, $\Cg$, is an integral over a function involving the gluon
self-energy and numerically estimated as $\Cg\approx 1.17201$ in the
dimensional regularization.  Here, $\mD$ is the gluon screening mass
induced by $\mu$, i.e., $\mD^2\equiv (2\alpha_s/\pi)\sum_f \mu_f^2$.  We
note that the bare quark masses in the hard loops are neglected
commonly in the HTL approximation (see Ref.~\cite{le_bellac_1996} for
a standard textbook).  The gluon sector is intact, so we just refer to
Refs.~\cite{Andersen:1999fw,*Andersen:1999sf, Haque:2014rua,
  Mogliacci:2013mca} for further details.

The quark part appears from the flavor-$f$ quark loop, i.e.,
$P_{\text{q},f}=\tr\ln G_f^{-1}$ where
$G_f^{-1}=\slashed{k}-M_f-\Sigma(k_0,\bk)$ and $k_0=i\tomega_n+\mu_f$
for flavor-$f$ quarks with $\tomega_n$ being the fermionic Matsubara
frequency.  For the self-energy expression, $\Sigma$, we need to
introduce the following notations according to
Refs.~\cite{Baier:1999db,Mogliacci:2013mca}, i.e.,
$A_0(k_0,k)\equiv k_0-(\mqf^2/k_0) \tTK(k_0,k)$, 
$A_s(k_0,k)\equiv k+(\mqf^2/k)[1-\tTK(k_0,k)]$, and the flavor-$f$ quark
screening mass is
$\mqf^2\equiv (\alpha_s/2\pi)(\Nc^2-1)/(2\Nc)\, \mu_f^2$.  The fermionic
HTLpt function in $d=3-2\epsilon$ spatial dimensions is:
\begin{equation}
  \tTK(k_0,k) = {}_2F_1 \biggl( \frac{1}{2},1; \frac{3}{2}-\epsilon;
  \frac{k^2}{k_0^2} \biggr)\,.
\end{equation}
Then, the self-energy for flavor-$f$ quarks is expressed as
$\slashed{k}-\Sigma(k_0,k)=A_0(k_0,k)\gamma^0
-A_s(k_0,k)\boldsymbol{\gamma}\cdot\hat{\bk}$.
In this work, we neglect the bare quark mass dependence in
$\Sigma(k_0,k)$; this treatment can be justified under the HDL
approximation.
In principle this effect can be taken into account by using the
effective action presented, e.g., in Ref.~\cite{Braaten:1992jj}.
The expression will, however, be extremely complicated, so we will
simply neglect here.
Nevertheless, it is unlikely that the bare quark mass plays an
important role for our main results, i.e., the reduction of the scale
dependent uncertainty.

The paramount advance in this work is the inclusion of bare mass
$M_f$, and the quark pressure deviates from
Refs.~\cite{Baier:1999db,Mogliacci:2013mca}.  Let us first write down
our final expression and then explain the notations next.  In the
flavor-$f$ quark sector the pressure contribution reads:
\begin{equation}
  P_{\text{q},f} = \mqf^4 \biggl[ \Cq(\eta_f)+\Dq(\eta_f)\ln\frac{\bLambda}{\mqf}
  \biggr] + P_{\text{qp},f} + P_{\text{Ld},f}\,.
  \label{eq:Pqf}
\end{equation}
We introduced $\Cq$ and $\Dq$ as functions of
$\eta_f \equiv 1 + M_f^2/(2\mqf^2)$.  These definitions involve the
following functions:
\begin{align}
  f_\pm(\bomega,\eta_f)
  & = \frac{\eta_f \pm \eta'(\bomega,\eta_f)}{1+\bomega^2}\,,
  \label{eq:f}\\
  \eta'(\bomega,\eta_f)
  &= \sqrt{\eta_f^2 \!-\! (1\!+\!\bomega^2) \biggl[
    \bigl(1\!-\!\tTK(i\bomega,1)\bigr)^2 \!+\!
    \frac{\tTK^2(i\bomega,1)}{\bomega^2}\biggr]}\,,
    \label{eq:eta}
\end{align}
where $\bomega$ is a dimensionless and continuous variable.  Then,
$\Cq$ and $\Dq$ are given by
\begin{align}
  \Cq(\eta_f)
  &= \sum_{\chi=\pm} \frac{1}{4\pi^3} \int_0^\infty d\bomega\,
    \biggl( f_\chi^2 \ln f_\chi - \frac{\partial f^2_\chi}
    {\partial\epsilon} \biggr) \notag\\
  &\qquad + \biggl(\frac{5}{4} - \ln 2\biggr) \Dq(\eta_f)\,,\label{eq:Cq}\\
  \Dq(\eta_f)
  &= -\sum_{\chi=\pm} \frac{1}{2\pi^3}\int_0^\infty d\bomega\,
    f_\chi^2\notag\\
  &=-\frac1{2\pi^2}(\eta_f^2 - 1)\,.
    \label{eq:Dq}
\end{align}
We note that $\Dq(\eta_f\to 1)\to 0$ and
$\Cq(\eta_f\to 1)\approx -0.03653$ as is consistent with
Ref.~\cite{Mogliacci:2013mca}.

The next term, $P_{\text{qp},f}$, in Eq.~\eqref{eq:Pqf} is the
quasi-particle contribution given by
\begin{equation}
  P_{\text{qp},f} = \frac{1}{\pi^2}\!\int_0^\infty \!\!\!\!dk\, k^2\!\!\!
  \sum_{\chi=\pm 1} \!\!\bigl[ (\mu_f-\omega_{f\chi})
  \theta(\mu_f-\omega_{f\chi}) \bigr]
  - \frac{\mu_f^4}{12\pi^2}\,.
  \label{eq:Pqpf}
\end{equation}
We note that the ideal term $\propto\mu_f^4$ is subtracted in the
above expression since we doubly pick up two pole contributions at
$\omega_{f\pm}$.  In Ref.~\cite{Baier:1999db} the quasi-particle
contribution was defined by taking the $\mqf^2$
derivative/integration, so that only the difference from the ideal
term was considered by construction, and the ideal term was not
subtracted but added.  Here, the quasi-particle poles,
$\omega_{f\pm}$, are solutions of the following implicit equations,
i.e.,
\begin{equation}
  0 = \omega_{f\pm} - \frac{\mqf^2}{k} Q_0\left(\frac{\omega_{f\pm}}{k}\right)
  \mp \sqrt{ M_f^2 + \left[
      k - \frac{\mqf^2}{k} Q_1\left(\frac{\omega_{f\pm}}{k}\right) \right]^2}
  \label{eq:disp}
\end{equation}
with $Q_0(x)\equiv (1/2)\ln[(x+1)/(x-1)]$ and
$Q_1(x)\equiv xQ_0(x) - 1$ being the Legendre functions.
Finally, the last term in
Eq.~\eqref{eq:Pqf} represents the contribution from the Landau
damping, which reads:
\begin{equation}
  P_{\text{Ld},f} = -\frac{1}{\pi^3} \int_0^{\mu_f} d\omega
  \int_{\omega}^\infty dk\, k^2\, \theta_{\text{q}f}(\omega,k;
  M_f,\mqf^2)\,.
  \label{eq:PLdf}
\end{equation}
The integrand is given by $\tan\theta_{\text{q}f} =
\mathcal{Y}/\mathcal{X}$ where
\begin{align}
  \mathcal{X}
  &= k^2-\omega^2 + M_f^2 + 2\mqf^2
    + \frac{\mqf^4}{k^2} \Biggl\{
    1-\frac{2\omega}{k}Q_0(k/\omega) \notag\\
  &\qquad\qquad\qquad - \frac{k^2-\omega^2}{k^2}
    \biggl[Q_0^2(k/\omega)-\frac{\pi^2}{4}\biggr] \Biggr\}\,, \label{eq:X}\\
  \mathcal{Y}
  &= \frac{\pi\mqf^4}{k^2} \biggl[
    \frac{\omega}{k} + \frac{k^2-\omega^2}{k^2}
    Q_0(k/\omega) \biggr]\,. 
    \label{eq:Y}
\end{align}
In this case $k\ge \omega$ holds and the argument of $Q_0$ should be
$k/\omega$, not $\omega/k$.  We also note that the subtraction at
finite $M_f$ is mass dependent, i.e.,
$\Delta P_{\text{q}}=\mqf^4 \Dq(\eta_f)/(2\epsilon)$.

For numerical calculations, we took
$M_u=M_d=0$ and $M_s(2\text{GeV})=100\,\text{MeV}$.  For $\Nc$ and
$\Nc$ in $\alpha(\bLambda)$ and $M_s(\bLambda)$ we took $\Nc=\Nf=3$.
This completes the explanation of the formulation necessary to draw
Fig.~\ref{fig:pe3}.


\begin{figure*}
  \centering
  \includegraphics[width=\columnwidth]{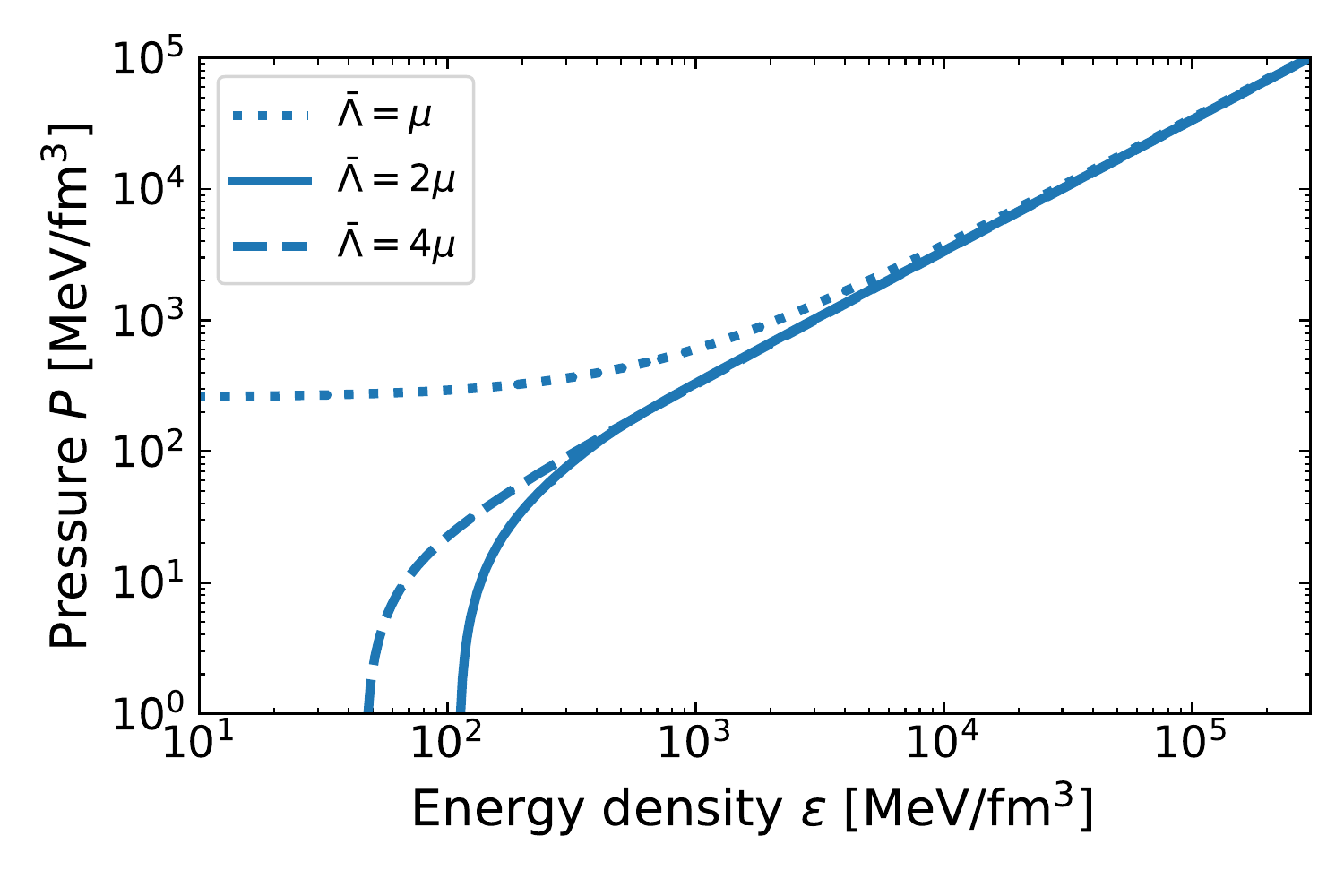}
  \includegraphics[width=\columnwidth]{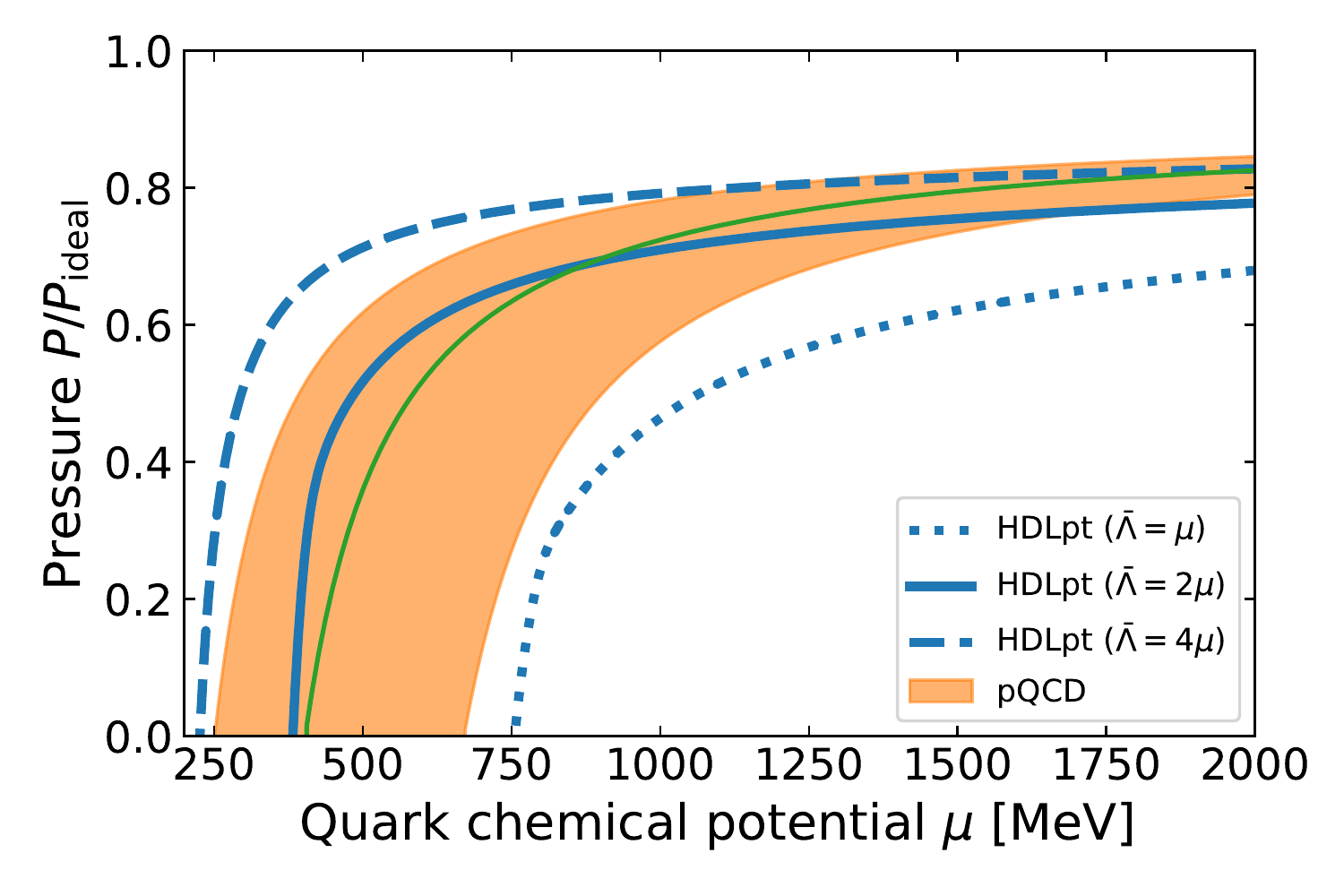}
  \caption{(Left) The same as Fig.~\ref{fig:pe3} in the previous section with
    an extended region of the energy density. (Right) The EoS
    expressed in the form of $P(\mu)$.}
  \label{fig:pemu}
\end{figure*}

In Fig.~\ref{fig:pemu}, we show the EoSs
calculated based on the formulation presented above.
In Fig.~\ref{fig:pemu} (Left), we show the EoS in the form of
$P(\varepsilon)$. This is the same plot as
Fig.~\ref{fig:pe3} above, but with an extended region of the energy
density.
Because of the uncertainty out of control at lower energy density it
is reasonable to truncate the plot around $\varepsilon \simeq
500\;\text{MeV}/\text{fm}^3$.

In Fig.~\ref{fig:pemu} (Right), we show the EoS in the form of $P(\mu)$.  It is
evident that the scale variation uncertainty in HDLpt is not small as
compared with the pQCD results.
Therefore, it is a quite nontrivial discovery that the scale variation
uncertainty in $P(\nB)$ is significantly smaller than that in
$P(\mu)$.

\begin{widetext}

\section{Details of integration: the quark contribution to the
  pressure}
\label{sec:details}

Here, we will elaborate the details of integration that
appears in the derivation of Eq.~\eqref{eq:Pqf} in the previous section.
The quark part of the pressure appears from the flavor-$f$ quark loop:
\begin{align}
  P_{\text{q},f}(T,\mu_f)
  &=\tr\ln G_f^{-1}\label{eq:trln} \\
  &= {\SumInt}_{\{K\}} \ln \det\left[\slashed{k} - M_f - \Sigma(i\tomega_n + \mu_f, k)\right] \notag\\
  &= 2 {\SumInt}_{\{K\}} \ln\left[A_S^2(i\tomega_n + \mu_f, k) + M_f^2
    - A_0^2(i\tomega_n + \mu_f, k)\right]\,,
    \label{eq:Pf}
\end{align}
where we write the sum-integral as ${\SumInt}_{\{K\}} =
T\sum_{\tomega_n} \int_{\bk}$ in $d=3-2\epsilon$ spatial dimensions for the momentum
integration.
The functions $A_0$ and $A_S$ are defined above.
We note that
$P_{\text{q},f}$ in Eq.~\eqref{eq:trln} can be regarded as
a leading contribution in the 2PI or the Cornwall-Jackiw-Tomboulis (CJT)
formalism~\cite{Cornwall:1974vz,Blaizot:2000fc}.
This explains why Eq.~\eqref{eq:trln} misses
an additional term, $\tr\Sigma G_f$, that may be
responsible for the deviation of $\mathcal{O}(\alpha_s)$, which will
be studied below.

\begin{figure}
  \centering
  \includegraphics[width=.47\textwidth]{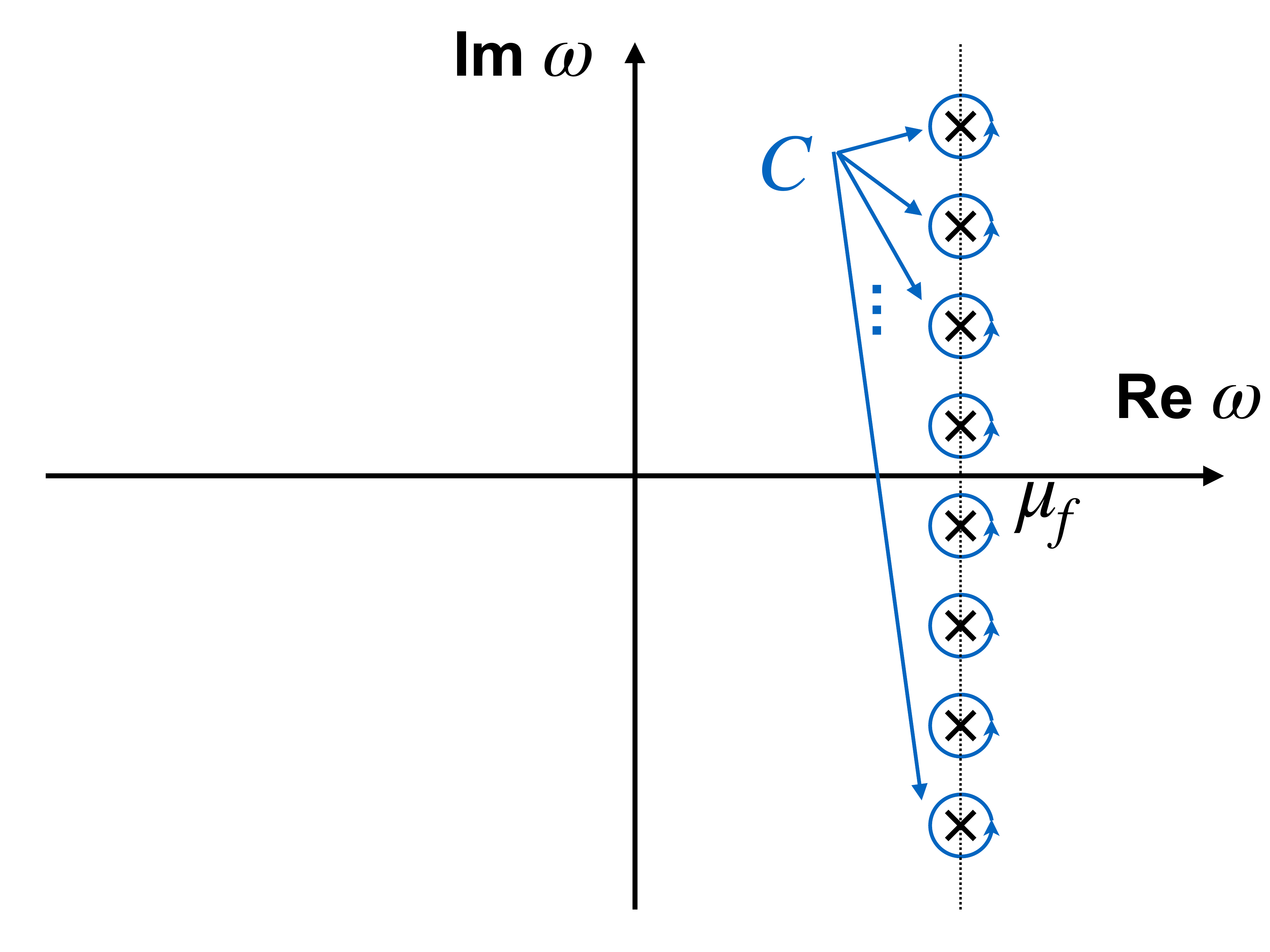}
  \hspace{1em}
  \includegraphics[width=.47\textwidth]{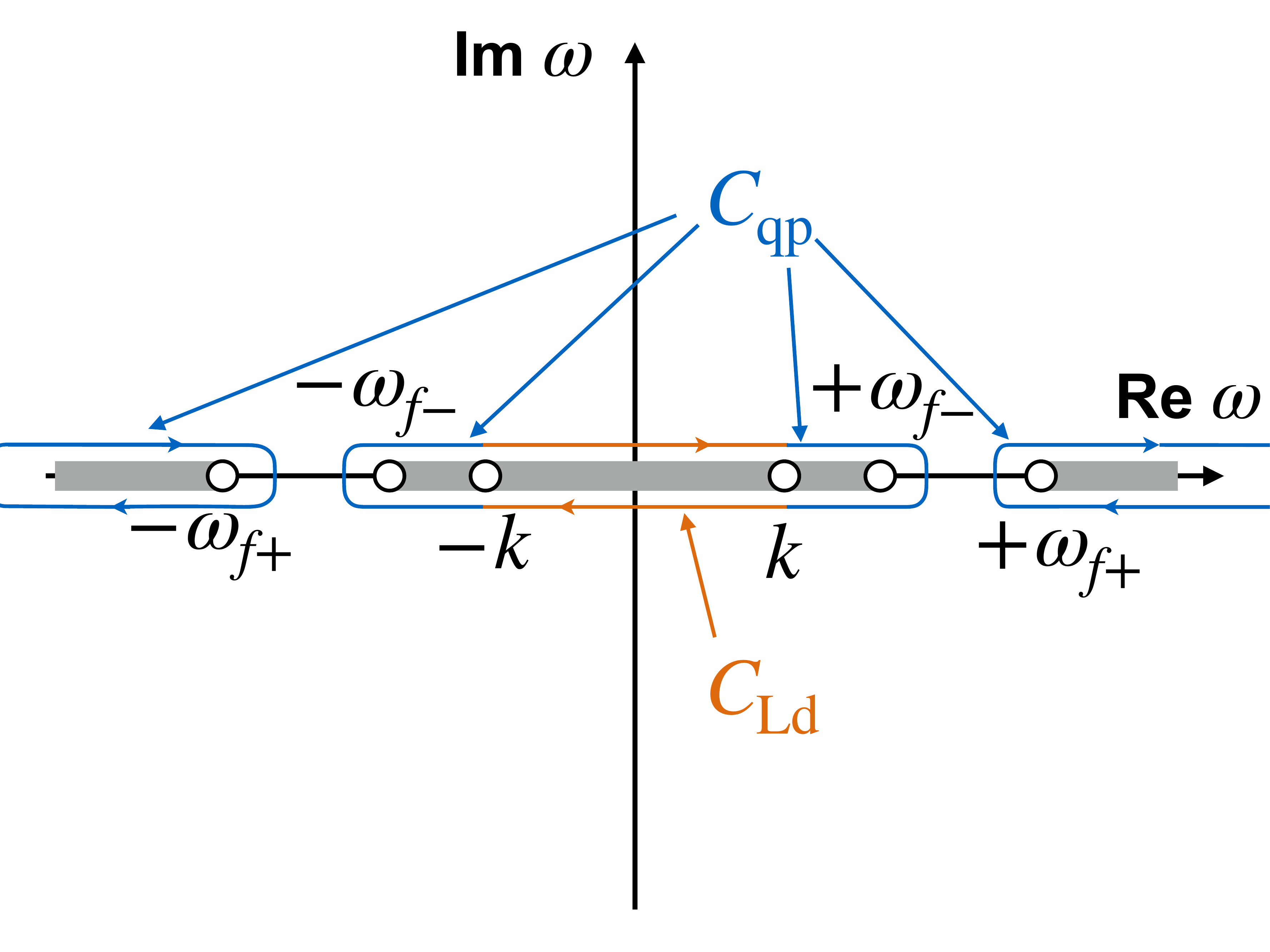}
  \caption{(Left) Original contour $\mathcal{C}$ corresponding to the Matsubara sum.  (Right)
    Deformed contours, $\mathcal{C}_{\rm qp}$ and $\mathcal{C}_{\rm Ld}$.}
  \label{fig:branch_qf}
\end{figure}

We recast the Matsubara sum into the contour integral
along $\mathcal{C}$ as depicted in the left
panel of Fig.~\ref{fig:branch_qf}.
We can deform the contour $\mathcal{C}$ into
$\mathcal{C}_{\rm qp} \cup \mathcal{C}_{\rm Ld}$, see
the right panel of Fig.~\ref{fig:branch_qf}.
We identify the terms from $\mathcal{C}_{\rm qp}$
and $\mathcal{C}_{\rm Ld}$ with the quasiparticle contribution and the
Landau damping contribution, respectively, according to
Refs.~\cite{Andersen:1999va, Mogliacci:2013mca}:
\begin{align}
  P_{\text{qp/Ld}, f}(T,\mu_f) =& \int_{\bk} \oint_{\mathcal{C}_{\rm qp/Ld}}
  \frac{d\omega}{2\pi i} \ln\left[A_s^2 (\omega, k) + M_f^2 -
    A_0^2(\omega, k)\right]
                                  \tanh\left(\frac{\beta(\omega-\mu_f)}{2}\right)\,.
\end{align}

The quasiparticle contribution to the integral (see the right panel of
Fig.~\ref{fig:branch_qf}) is
\begin{align}
  P_{\text{qp}, f}
  =& \int_{\bk} \left\{ \int_{\tilde{\omega}_{f+}}^\infty \frac{d\omega}{2\pi}
  \left[\Disc \arg\left(A_s^2 (\omega, k) + M_f^2 - A_0^2(\omega,
    k)\right)\right]
  \left[\tanh\left(\frac{\beta(\omega-\mu_f)}{2}\right) -
    \tanh\left(\frac{\beta(-\omega-\mu_f)}{2}\right)\right] \right. \notag\\
  &\qquad+  \left.\int_k^{\tilde{\omega}_{f-}} \frac{d\omega}{2\pi}
  \left[\Disc \arg\left(A_s^2 (\omega, k) + M_f^2 - A_0^2(\omega,
    k)\right)\right]
  \left[\tanh\left(\frac{\beta(\omega-\mu_f)}{2}\right) -
    \tanh\left(\frac{\beta(-\omega-\mu_f)}{2}\right)\right] \right\} \notag\\
  =& \int_{\bk} \left\{ \int_{\tilde{\omega}_{f+}}^\infty \frac{d\omega}{2\pi}
  \left(-2\pi\right)
  \left[2 \!-\! \frac{2}{e^{\beta(\omega-\mu_f)} + 1} \!-\!
    \frac{2}{e^{\beta(\omega+\mu_f)} + 1}\right]\right. \notag 
                       + \left. \int_k^{\tilde{\omega}_{f-}} \frac{d\omega}{2\pi}
  \left(2\pi\right)
  \left[2 \!-\! \frac{2}{e^{\beta(\omega-\mu_f)} + 1} \!-\!
    \frac{2}{e^{\beta(\omega+\mu_f)} + 1}\right] \right\} \notag\\
  =& 2\int_{\bk} \sum_{\chi,s=\pm} T \ln\left[1 + e^{-\beta\left(\omega_{f\chi} + s \mu_f \right)}\right]
  - 2\int_{\bk} \sum_{s=\pm} T\ln\left[1+e^{-\beta\left(k +s \mu_f \right)}\right]
  + 2\int_{\bk}[\omega_{f+}(k) + \omega_{f-}(k) - k]\,,
  \label{eq:Pqsqp}
\end{align}
where we defined as $\Disc f(\omega) \equiv f(\omega + i0^+) -
f(\omega - i0^+)$,
used
$A_{0, s}^2(\omega,k) =A_{0, s}^2(-\omega,k)$,
and dropped an irrelevant infinity from the upper bound of the
$\omega$-integration.
The dispersion relation for quarks $\omega_{f\pm}$ is
obtained by solving Eq.~\eqref{eq:disp} above.
For the moment we can drop the third term in Eq.~\eqref{eq:Pqsqp} that is
independent of $T$ and $\mu_f$ (which will be reassembled later).
Finally, we obtain:
\begin{equation}
  P_{\text{qp},f}(T=0, \mu_f) = \frac{1}{\pi^2}\!\int_0^\infty \!\!\!\!dk\, k^2\!\!\!
  \sum_{\chi=\pm 1} \!\!\bigl[ (\mu_f-\omega_{f\chi})
  \theta(\mu_f-\omega_{f\chi}) \bigr]
  - \frac{\mu_f^4}{12\pi^2}\,,
\end{equation}
which completes the derivation of Eq.~\eqref{eq:Pqpf} above.
The $s=-1$ term in the sum of Eq.~\eqref{eq:Pqsqp}
vanishes at $T\to 0$ because of the step
function $\theta(-\mu_f - \omega_{f\chi})$.

The Landau damping contribution to the integral is
\begin{align}
  P_{\text{Ld}, f}
  =& \int_{\bk} \int_{-k}^k \frac{d\omega}{2\pi}
  \Disc \arg\left(A_s^2 (\omega, k) + M_f^2 - A_0^2(\omega,
    k)\right) \tanh\left(\frac{\beta(\omega-\mu_f)}{2}\right) \notag\\
  =& -\frac1{\pi} \int_{\bk} \int_0^k d\omega\,2\theta_{\text{q}f}(\omega,k;M_f^2,\mqf^2)
  \left[\frac{1}{e^{\beta(\omega-\mu_f)} + 1} +
    \frac{1}{e^{\beta(\omega+\mu_f)} + 1} - 1 \right]\,.
  \label{eq:PqsLd}
\end{align}
In the last line we introduced
[with $\mathcal{X}$ and $\mathcal{Y}$ defined in Eqs.~\eqref{eq:X} and
\eqref{eq:Y} above, respectively]:
\begin{align}
  &2\theta_{\text{q}f} = 2\arctan\mathcal{Y}/\mathcal{X} =
  \Disc \arg\left(A_s^2 (\omega, k) + M_f^2 - A_0^2(\omega,k)\right) \notag\\
  &= \Disc \arctan\left\{\frac{\Im\left[A_s^2 (\omega, k) + M_f^2 - A_0^2(\omega,
    k)\right]}{\Re\left[A_s^2 (\omega, k) + M_f^2 - A_0^2(\omega,
      k)\right]}\right\} \notag\\
  &= \Disc \arctan \left\{\frac{\frac{\mqf^4}{k^2} \left[ - 2
      \Im\left({}_2F_1(\tfrac12, 1; \tfrac32;
      \tfrac{k^2}{\omega^2})\right) - \frac{k^2 - \omega^2}{\omega^2} \Im
      \left({{}_2F_1(\tfrac12, 1; \tfrac32;
        \tfrac{k^2}{\omega^2})}^2\right) \right]}
  {k^2 - \omega^2 + M_f^2 + 2\mqf^2 + \frac{\mqf^4}{k^2} \left[1 - 2
      \Re\left({}_2F_1(\tfrac12, 1; \tfrac32;
      \tfrac{k^2}{\omega^2})\right) - \frac{k^2 - \omega^2}{\omega^2}
      \Re\left({{}_2F_1(\tfrac12, 1; \tfrac32;
        \tfrac{k^2}{\omega^2})}^2\right)\right]}\right\} \notag\\
  &= 2\arctan \left\{\frac{\frac{\mqf^4}{k^2} \left[ - 2
      \left(-\frac{\pi\omega}{2k}\right)
       - \frac{k^2 - \omega^2}{\omega^2}
       \left(-\frac{\pi\omega^2}{2k^2}\qzero{k}{\omega}\right)
       \right]}
  {k^2 - \omega^2 + M_f^2 + 2\mqf^2 + \frac{\mqf^4}{k^2} \left[1 - 2
      \frac{\omega}{2k}\qzero{k}{\omega} - \frac{k^2 - \omega^2}{\omega^2}
      \frac{\omega^2}{4k^2} \left[\qzero{k}{\omega}^2 -
        \pi^2\right]\right]}\right\}\,.
\end{align}
Again, we only keep the $T$ and $\mu_f$ dependent parts
in Eq.~\eqref{eq:PqsLd}, so that the $T\to 0$ limit leads to
\begin{equation}
  P_{\text{Ld},f}(T=0, \mu_f) = -\frac{1}{\pi^3} \int_0^{\mu_f} d\omega
  \int_{\omega}^\infty dk\, k^2\, \theta_{\text{q}f}(\omega,k;
  M_f,\mqf^2)\,,
\end{equation}
which completes the derivation of Eq.~\eqref{eq:PLdf} in the previous
section.

We here reassemble the $T$ and $\mu_f$ independent terms that
we dropped above.
To this end it is convenient to think of the $T=\mu_f=0$ limit in
Eq.~\eqref{eq:Pf}, in which the Matsubara sum reduces to
$T\sum_{n} \to \int_{-\infty}^\infty \frac{d\bomega}{2\pi}$,
so that the pressure reads:
\begin{align}
  P_{\text{q}f}^\star &= 2 \int_{-\infty}^\infty
  \frac{d\bomega}{2\pi} \int_{\bk}  \ln\left[A_{S}^2 (i\bomega,
    k) + M_f^2 - A_0^2(i\bomega, k)\right] \notag\\
  &= 4 \int_{0}^\infty \frac{d\bomega}{2\pi} \int_{\bk} k
  \ln\left\{(1 + \bomega^2) k^2 + M_f^2 + 2\mqf^2 +
  \frac{\mqf^4}{k^2}\left[\left(1 - \tTK(i\bomega,
    1)\right)^2 - \frac{\tTK^2(i\bomega, 1)}{\bomega^2}\right]\right\} \notag\\
  &= -\frac{\bLambda^{2\epsilon} e^{\gamma_{\rm E}
      \epsilon}}{4\pi^{5/2}} \frac{\Gamma(2 - \epsilon)
    \Gamma(\epsilon - 2)}{\Gamma(\tfrac32 - \epsilon)} \mqf^{4-2\epsilon}
  \int_0^\infty d\bomega \, \left[\left(f_{+}(\bomega,\eta_f)\right)^{2 -
      \epsilon} + \left(f_{-}(\bomega,\eta_f)\right)^{2 - \epsilon}\right]\,,
\end{align}
where we used the following integral:
\begin{equation}
  \int_0^\infty dk\, k^\alpha \ln(k^2 + m^2) = \frac{\Gamma\left(\frac{1+\alpha}2\right)\Gamma\left(\frac{1-\alpha}2\right)}{1+\alpha}m^{1+\alpha}\,.
\end{equation}
The function $f_\pm(\bomega,\eta_f)$ with
$\eta_f\equiv1+M_f^2/(2\mqf^2)$ is defined as in Eqs.~\eqref{eq:f} and
\eqref{eq:eta} above.
The limit of $\epsilon\to 0$ gives:
\begin{align}
  P_{\text{q}f}^\star
  =& -\frac{\mqf^4}{4\pi^3} \left(\frac1{\epsilon} +
  \ln\frac{\bLambda^2}{\mD^2} + \frac52 - 2\ln2 \right)
  \left[\sum_{\chi=\pm}\int_0^\infty d\bomega \, f_{\chi}^2 -\epsilon
    \sum_{\chi=\pm}\int_0^\infty d\bomega \, \left(f_{\chi}^2 \ln
    f_{\chi} - 2 f_{\chi} \frac{\partial f_{\chi}}{\partial
      \epsilon}\right)\right] \notag\\
  =& \mqf^4 \biggl[ \Cq(\eta_f)+\Dq(\eta_f)\ln\frac{\bLambda}{\mqf}
    \biggr] + \mqf^4 \Dq(\eta_f) \frac1{2\epsilon}\,.
\end{align}
The constants $\Cq$ and $\Dq$ are defined in Eqs.~\eqref{eq:Cq} and
\eqref{eq:Dq} in the previous section, respectively.
The ultraviolet divergence is subtracted by the term $\Delta P_{\rm q}$
in Eq.~\eqref{eq:PHDL} above:
\begin{align}
  \Delta P_{\rm q} = \mqf^4 \Dq(\eta_f) \frac1{2\epsilon}\,.
\end{align}
In this way the above procedures complete the derivation of
Eq.~\eqref{eq:Pqf} in the previous section.

\end{widetext}

\section{Discussions}
\label{sec:discussions}

Here, we discuss the speed of sound that could exceed the conformal
limit, and the robustness against the $\mathcal{O}(\alpha_s)$
corrections to match the conventional pQCD calculation.

\subsection{Speed of sound}
\begin{figure}
  \includegraphics[width=\columnwidth]{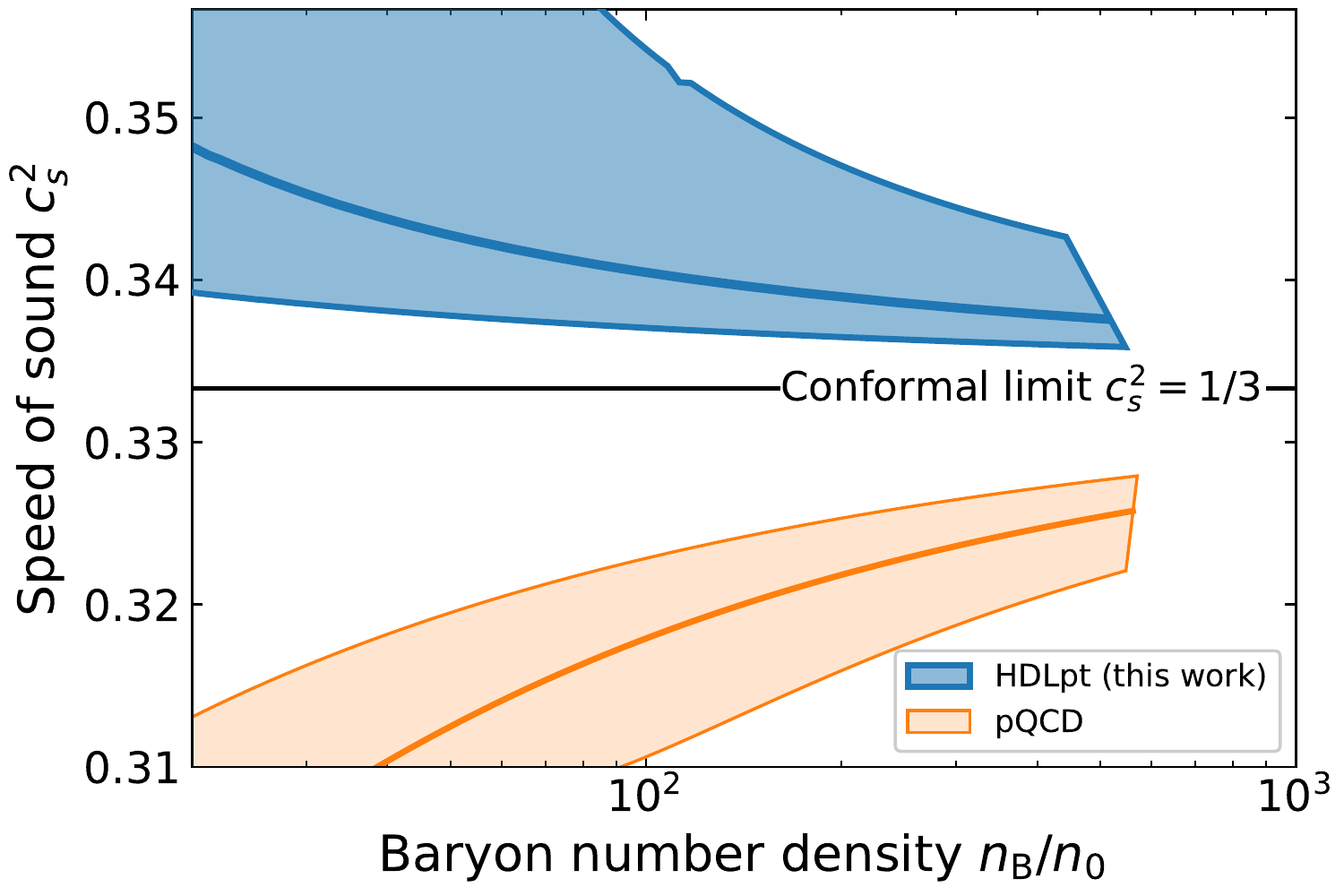}
  \caption{Speed of sound $c_s^2$ from the EoSs;  the blue band
    represent the results from our HDLpt EoS, and the orange band from
    the pQCD for reference.}
  \label{fig:cs2nb}
\end{figure}

The EoS from our resummed perturbation theory has a notable feature in
addition to the smaller uncertainty.  We have calculated the speed of
sound, $c_s^2=\partial P/\partial \varepsilon$, which is depicted in
Fig.~\ref{fig:cs2nb}.  To make clear the relevance to the neutron star
environment, we chose the horizontal axis as the baryon number density
$\nB$ in the unit of the normal nuclear density $n_0$.

There is an empirical conjecture to claim that the speed of sound may
not exceed the conformal limit, i.e., $c_s^2=1/3$.  In the high
density limit, asymptotically, all mass scales and interactions are
negligible and the conformal limit should be eventually saturated.  In
the pQCD calculation, the first correction from the conformal limit is
negative, so that the conformal limit is approached from $c_s^2<1/3$
with increasing density.  Also at finite temperature, the lattice-QCD results
demonstrate that the conformal bound $c_s^2<1/3$
holds~\cite{Borsanyi:2013bia,*Bazavov:2014pvz}.  Known examples of
QCD calculations seem to respect the conformal limit (see
Ref.~\cite{Son:2000xc} for an exception at finite isospin chemical
potential).  However, no field-theoretical proof exists to guarantee
$c_s^2 < 1/3$.  The recent analysis based on neutron star data,
especially the two-solar-mass condition, indeed suggest a possibility
of $c_s^2 > 1/3$ at sufficiently high baryon
density~\cite{Bedaque:2014sqa,Tews:2018kmu,Drischler:2020fvz}.

Figure~\ref{fig:cs2nb} shows that our resummed EoS slightly violates
the conformal bound and $c_s^2$ approaches $1/3$ from above.
It is evident that our result is a counterexample to the conjecture of
$c_s^2 < 1/3$.  The quantitative difference is numerically small
between EoSs from our HDLpt and pQCD, and the violation of the
conformal bound is tiny, but this comparison on Fig.~\ref{fig:cs2nb}
implies that one should be careful about the robustness of the speed
of sound bound (see, for example, discussions in
Refs.~\cite{Annala:2019puf,*Annala:2019eax}).

\subsection{$\mathcal{O}(\alpha_s)$ correction}

\begin{figure}
  \centering
  \includegraphics[width=\columnwidth]{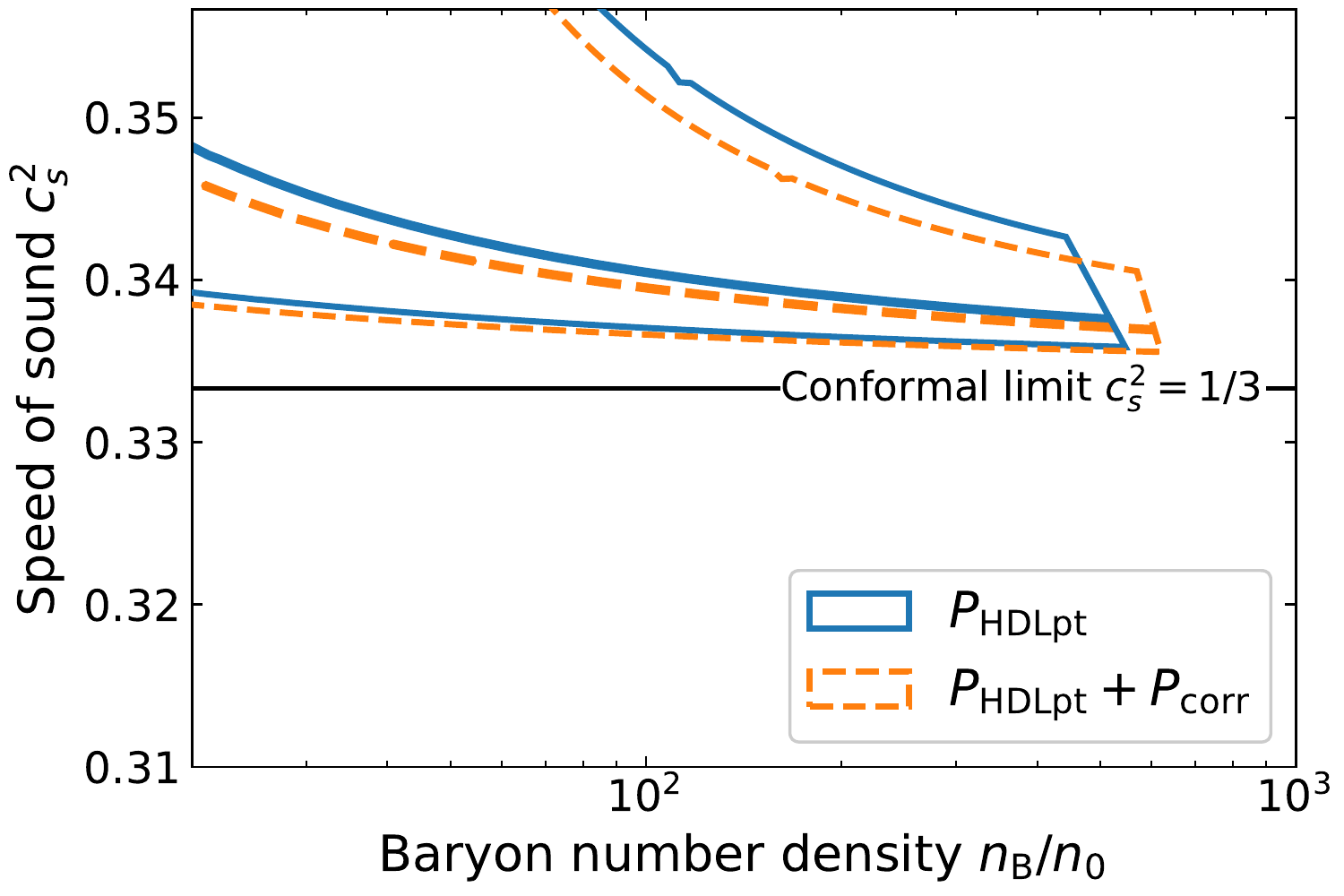}
  \caption{Comparison between the speed of sound evaluated from
    $\PHDL$ and $\PHDL + P_{\text{corr}}$.}
    \label{fig:cs2_corr}
\end{figure}

The HDLpt has a deviation of $\mathcal{O}(\alpha_s)$ in the pressure,
as mentioned in the beginning, from the conventional pQCD calculation.
Our HDLpt predicts $c_s^2 > 1/3$ even if we add a correction
to match the $\mathcal{O}(\alpha_s)$ terms.  For analytical simplicity
we will show the calculation in the massless case only.
It is known that the expansion of $P_{\text{HDLpt}}$ in powers of
$\mqf/\mu_f \ll 1$ gives, for $\Nc=3$~\cite{Baier:1999db}:
\begin{equation}
  \frac{\PHDL}{P_{\text{ideal}}} \approx 1 - 6
  \frac{\mqf^2}{\mu_f^2} +
  \mathcal{O}\left(\frac{\mqf^4}{\mu_f^4}\right) = 1 -
  4\frac{\alpha_s}{\pi} + \mathcal{O}(\alpha_s^2)\,,
\end{equation}
where the ideal pressure is $P_{\text{ideal}} =
\Nc\Nf\mu_f^4/(12\pi^2)$.
The conventional pQCD result
is~\cite{Freedman:1976xs,*Freedman:1976dm,*Freedman:1976ub,Baluni:1977ms}
\begin{equation}
  \frac{P_{\text{pQCD}}}{P_{\text{ideal}}} = 1 - 2\frac{\alpha_s}{\pi}
  + \mathcal{O}(\alpha_s^2)\,.
\end{equation}
Therefore we can match the $\mathcal{O}(\alpha_s)$ terms by adding the
following correction to $\PHDL$:
\begin{equation}
  P_{\text{corr}} = 2\frac{\alpha_s}{\pi}P_{\text{ideal}}\,.
\end{equation}

In Fig.~\ref{fig:cs2_corr} we plot the speed of sound evaluated by
$\PHDL$ and $\PHDL + P_{\text{corr}}$ both in the massless case.
The figure~\ref{fig:cs2_corr} clearly shows that even with the
$P_{\text{corr}}$ correction, the speed of sound still approaches
$c_s^2 = 1/3$ from the above as the density increases.
This implies that $c_s^2 > 1/3$ could be attributed to the
higher order effects from the resummation.

\section{Summary}
\label{sec:summary}
In this work we showed results with the
smaller scale variation uncertainty for the
cold dense matter EoS in the form of $P(\varepsilon)$.  The formalism
we adopted here is the HDLpt, which has already been successful in
finite temperature QCD{}.  The important observation is that, as
compared to the pQCD calculation, quarks are screened by self-energy
insertions, and the baryon density is suppressed.  This means that the
corresponding chemical potential for a given baryon density is shifted
to be larger, particularly for $\bLambda = \mu$.  It was the source of
the large uncertainty in the pQCD calculation, so the improvements for
$\bLambda = \mu$ helps lessen the uncertainty band.  We also emphasize
the importance of the inclusion of a bare quark mass and we
numerically solved the $\beta$ equilibrium and charge neutrality
conditions.  Our treatments with the bare quark mass are messy, but
contributions from finite strange quark mass are crucial for the
realistic environments of neutron stars under the $\beta$ equilibrium.
Our results constitute a QCD-based example of the conformal limit
violation at finite density, which can be in consonance with the
state-of-the-art neutron star observations.  It would be an exciting
program to apply our EoS to the neutron star phenomenology.  We will
report phenomenological implications soon.

\begin{acknowledgments}
The authors thank
Yuya~Abe
for useful discussions.
This work was supported by Japan Society for the Promotion of Science
(JSPS) KAKENHI Grant Nos.\ 18H01211, 19K21874 (KF) and 20J10506 (YF).
\end{acknowledgments}

\bibliography{HTLpt}
\bibliographystyle{apsrev4-2}

\end{document}